\documentclass[twocolumn,showpacs,prl]{revtex4}
\usepackage{dcolumn}
\usepackage{graphicx}
\usepackage{graphicx,amssymb,amsmath,revsymb}
\usepackage{natbib}
\usepackage{multirow}

\begin{document}

\title{Large--scale structure from superdiffusion in a driven dissipative system}

\author{David A. Head$^{1,2}$}
\author{Hajime Tanaka$^{1}$}

\affiliation{$^{1}$Institute of Industrial Science, University of Tokyo, Meguro-ku, Tokyo 153-8505, Japan}
\affiliation{$^{2}$Institut f\"ur Festk\"orperforschung, Theorie II, Forschungszentrum J\"ulich 52425, Germany}

\date{\today}

\begin{abstract}
A system far from equilibrium is characterized by unconventional many--body dynamical effects, which can lead to anomalous density fluctuations and mass transport. Interestingly, these structural and dynamic features often emerge simultaneously in driven dissipative systems. Here we seek an origin of their co-existence by numerical simulations of a two-dimensional driven granular gas. We reveal a causal link between superdiffusive transport and giant density fluctuations. The kinetic dissipation upon particle collisions depends on the relative velocity of colliding particles, and is responsible for the self-generated large-scale persistent directional motion of particles that underlies the link between structure and transport. This scenario is supported by a simple scaling argument. 
\end{abstract}



\maketitle

%
%

Many materials of biological or industrial importance are driven far from thermodynamic equilibrium by an imposed energy flux, mediated by {\em e.g.} motor proteins in living cells, or boundary--induced flow of fluids or particulate matter~\cite{Mukamel,GranularReview,ActiveGel,DustyPlasma,IntracellularTransport,Majda1999,SelfPropelled,Lechenault,Heussinger2010}. Given a mechanism for energy dissipation and sufficient relaxation time, such systems may reach a statistical steady state in which macroscopic quantities remain constant. Unlike thermodynamic equilibrium, however, the underlying sequence of micro--states admits cyclical currents~\cite{Mukamel}, and consequently driven dissipative systems exhibit a richer variety of structure formation than thermal systems~\cite{GranularReview,ActiveGel}. The transport of matter within such systems can also be anomalous, such as when the mean--squared particle displacements grow faster than linearly in time. Examples of such {\em superdiffusive} systems include dusty plasmas~\cite{DustyPlasma}, intracellular transport~\cite{IntracellularTransport}, turbulent fluids~\cite{Majda1999}, self--propelled particles~\cite{SelfPropelled} and granular media~\cite{Lechenault,Heussinger2010}.

Interparticle interactions, even if just steric hinderance, inevitably lead to some form of correlation in the motion of nearby particles. It is therefore expected that superdiffusion should have a measurable many--body consequence and a corresponding spatial signature. There is some suggestion of this in the aforementioned materials: Superdiffusion and a diverging dynamic correlation length have been observed in both experiments on sheared frictional granular media~\cite{Lechenault} and simulations of frictionless particles~\cite{Heussinger2010} near the `jamming' transition dividing rigid from non--rigid packings; and hydrodynamic equations for self--propelled particles admit superdiffusion and long--range ordering~\cite{SelfPropelled}. These examples suggest superdiffusion can be linked with long range spatial correlations, but it is not clear for which systems, if any, the link is causal.

Here we investigate the link between anomalous mass transport and large scale structure formation in simulations of a model driven dissipative system, in which particles are uniformly agitated on the single--particle level and dissipate energy through short--range pair interactions. This flux produces superdiffusive particle transport that can be attributed to the spontaneous formation of convective currents~\cite{Majda1999}. We provide a simple theory predicting long--range structure formation as a consequence of the anomalous transport, and confirm the expected small wavenumber divergence in the static structure factor. As an intermediate calculation, we predict and observe {\em giant number fluctuations} in which variations in the particle number $N$ exceed the classical $N^{1/2}$ expectation. This phenomenon was previously predicted for active nematics and observed in vibrated granular rods~\cite{GNF}, and also claimed for spherical particle monolayers~\cite{AransonNarayan}, but failure to reach ergodicity cast doubt upon this latter case. We confirm both ergodicity and convergence with system size, providing unequivocal evidence that giant fluctuations can exist in non--equilibrium systems of isotropic particles.

%
%

{\em Model.}---Our system, inspired by vibrated granular monolayers~\cite{GranularReview}, consists of radially symmetric particles with short--range repulsive, dissipative interactions, driven by a homogeneous and isotropic, Langevin--like force noise. The particles are discs $\alpha$ with polydisperse diameters $d^{\alpha}$ and equal mass density. Two discs $\alpha$ and $\beta$ interact with equal--and--opposite forces when their centers are separated by a distance $R^{\alpha\beta}<\frac{1}{2}(d^{\alpha}+d^{\beta})$. The interaction has a repulsive conservative component of magnitude $f^{\rm cons}=\mu[1-2R^{\alpha\beta}/(d^{\alpha}+d^{\beta})]$ acting along the line of centers, and a dissipative term ${\bf f}^{\rm diss}=\eta({\bf v}^{\alpha}-{\bf v}^{\beta})$ which reduces the relative velocity. These interactions conserve momentum but dissipate (kinetic) energy. The driving term consists of a spatio--temporally uncorrelated fluctuating Gaussian force field $\boldsymbol\xi({\bf r},t)$ that obeys white noise statistics, $\langle\xi_{i}({\bf r}_{1},t_{1})\xi_{j}({\bf r}_{2},t_{2})\rangle=\Gamma\delta_{ij}\delta({\bf r}_{1}-{\bf r}_{2})\delta(t_{1}-t_{2})$. This field does not conserve momentum locally, but it is imposed globally to ensure a fixed system centre of mass. Note that  there is no frictional term with an implicit solvent or substrate.

Discs with diameters uniformly distributed over the range $[0.7\langle d\rangle,1.3\langle d\rangle]$ are randomly placed in an \mbox{$L\times L$} simulation cell with periodic boundaries, until the required area fraction $\phi=L^{-2}\sum_{\alpha}\pi(d^{\alpha}/2)^{2}$ has been achieved. Here we consider only the case $\phi=0.5$, corresponding to intermediate densities far below the jamming transition for this polydispersity, $\phi_{\rm J}\approx0.843$~\cite{Head2009}. Rather than fix the force noise $\Gamma$, we instead choose $\Gamma$ to give the required mean kinetic energy $K(t)=\sum_{\rm a}\frac{1}{2}m^{\alpha}[v^{\alpha}(t)]^{2}$ in the steady state, with $m^{\alpha}$ the mass of particle~$\alpha$. This $\Gamma$ was interpolated from data generated by a series of calibration runs for each~$\eta$.

To determine when a statistical steady state has been reached, we measure the two--time mean squared displacement \mbox{$\langle\Delta r^{2}(t_{\rm w},t_{\rm w}+t)\rangle=\langle|{\bf r}(t_{\rm w}+t)-{\bf r}(t_{\rm w})|^{2}\rangle$} for particle displacements between times $t_{\rm w}$ and $t_{\rm w}+t$. Stationarity is assumed when \mbox{$\langle\Delta r^{2}(t_{\rm w},t_{\rm w}+t)\rangle$} ceases to vary with~$t_{\rm w}$ and time translational invariance has been achieved, {\em i.e.} $\langle\Delta r^{2}(t_{\rm w},t_{\rm w}+t)\rangle\equiv\langle\Delta r^{2}(t)\rangle$. All quantities are expressed in dimensionless forms after suitable scaling by the bare distance $\langle d\rangle$, time $t_{0}=\sqrt{\langle d\rangle\langle m\rangle/\mu}$, dissipation coefficient $\eta_{0}=\sqrt{\mu\langle m\rangle/\langle d\rangle}$ and energy $K_{0}=\langle d\rangle\mu$. For this study, all 9 combinations of $K(\infty)/K_{0}=5\times10^{-3}$, $5\times10^{-4}$ and $5\times10^{-5}$ and $\eta/\eta_{0}\approx 0.04$, 0.08 and 0.24 were used. Snapshots are given in~\cite{EPAPS}.

\begin{figure}[htpb]
\center{\includegraphics[width=9cm]{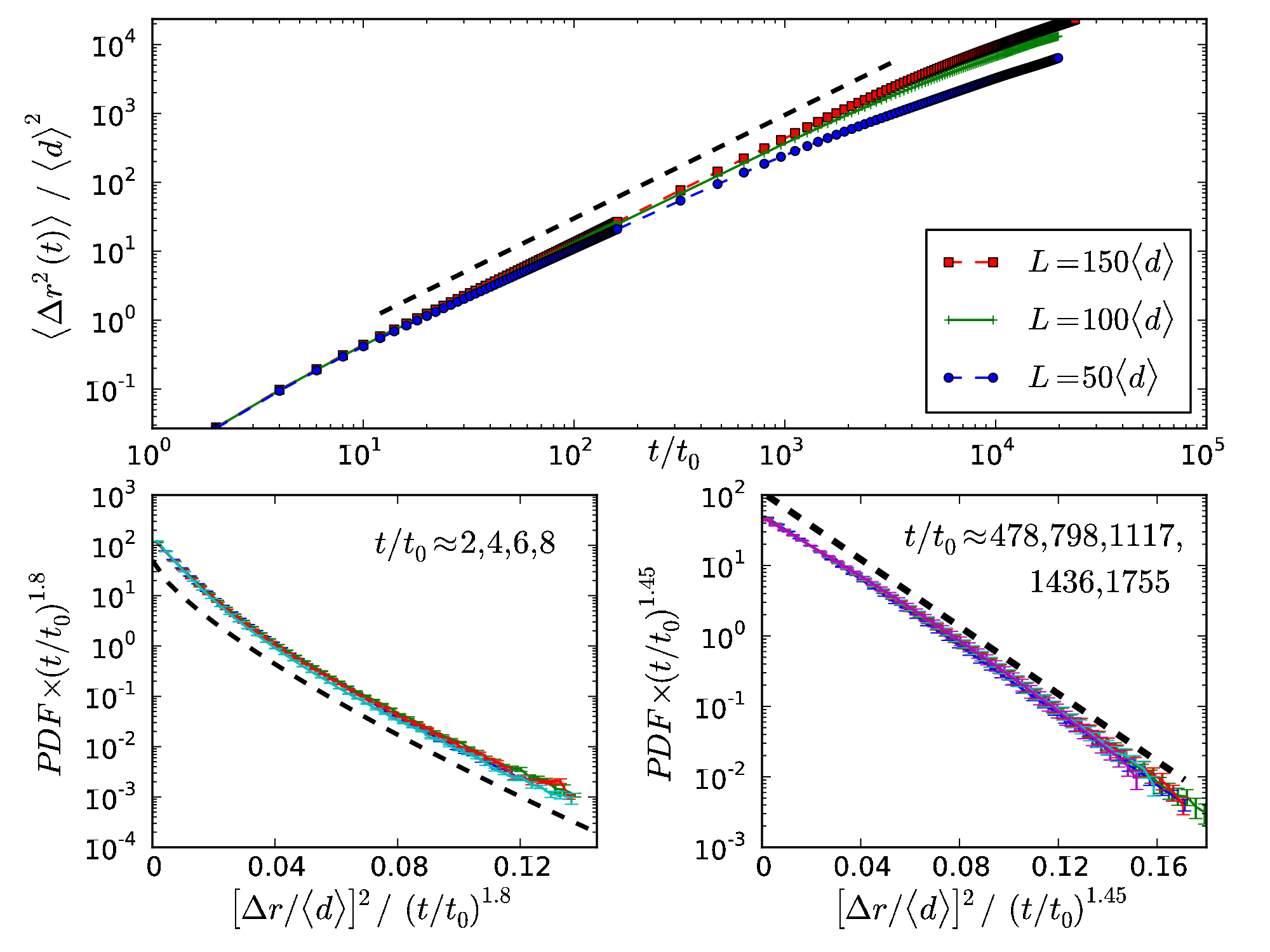}}
\caption{{\em (Color online)}
{\em (Top panel)}~Mean--squared displacement $\langle\Delta r^{2}(t)\rangle$ for the given system sizes~$L$, $K/K_{0}=5\times10^{-4}$ and $\eta/\eta_{0}=0.08$. The dashed line has slope~1.5.
{\em (Lower left panel)}~Probability distribution function (PDF) of displacements for $L=150\langle d\rangle$ at short times, scaled by $t^{1.8}$. The dashed line is $\propto \exp\{-A(\Delta r)^{1.5}\}$.
{\em (Lower right panel)}~Same for larger times, scaled by $t^{1.45}$. The dashed line is Gaussian.
}
\label{f:msd}
\end{figure}

%
%
{\em Results.}---For all $K$ and~$\eta$ investigated, the mean squared displacement exhibits 3 regimes (see Fig.~\ref{f:msd}): {\em (i)}~A rapid initial growth for trajectories shorter than the particle size, $\langle\Delta r^{2}(t)\rangle\ll\langle d\rangle^{2}$; {\em (ii)}~Superdiffusive motion at intermediate times, $\langle\Delta r^{2}(t)\rangle\sim t^{1+a}$ with $0<a<1$; {\em (iii)}~Normal diffusion $\langle\Delta r^{2}(t)\rangle\sim t$ at late times. This latter regime moves to later times for larger system sizes~$L$, and we infer it is a finite size effect and the true asymptotic behavior is the superdiffusive regime~{\em (ii)}. The crossover from {\em (i)} to {\em (ii)} is also evident in the distribution $P(\Delta r^{2},t)$ of squared particle displacements $\Delta r^{2}$ over the lag time~$t$. For each regime it is possible to collapse $P(\Delta r^{2},t)$ onto a single curve after scaling the axes by a power of~$t$ while preserving normalization, \mbox{$P(\Delta r^{2},t)=t^{-\nu}p(\Delta r^{2}/t^{\nu})$}, as shown in Fig.~\ref{f:msd}. For short times we consistently find $\nu$ close to~2, suggesting we are approaching the expected ballistic regime with $P(\Delta r^{2},t)\sim p^{\rm ball}([\Delta r/t]^{2})$ and \mbox{$\Delta r/t$} the velocity. The collapsed curve can be fitted to a `stretched' exponential \mbox{$p^{\rm ball}\sim\exp\{-A(\Delta r/t)^{3/2}\}$}, consistent with velocity distributions in granular gases~\cite{GranVelDist}. More relevant here are large times for which collapse is possible with $\nu\approx1.5$, but now the master curve is, to good approximation, a Gaussian with no fat tail. Superdiffusion is therefore identified with the anomalous broadening of the whole distribution, {\em i.e.}~the variance $\sigma^{2}(t)\sim t^{\nu}\sim t^{1+a}$, rather than large jumps by a small subpopulation of particles as observed in high density frictional packings~\cite{Lechenault}.

%
%

Fitted values of the exponents $a$ and $\nu$ for different $K$ and $\eta$ are given in Table~\ref{t:exps}. These values scatter around \mbox{$\nu=1+a=3/2$}, and while random errors for some points are not consistent with this value, we cannot rule out small systematic errors of $\approx0.1-0.2$ resulting from the limited scaling regimes used for fitting. We believe that the true asymptotic value is $a=1/2$ and is a consequence of the global driving and momentum conservation. To see this, first recall that $\langle \Delta r^{2}(t)\rangle$ is directly related to the velocity autocorrelation function $R(t)=\langle{\bf v}(0)\cdot{\bf v}(t)\rangle$, where ${\bf v}(t)$ refers to the velocity of a tagged particle at time~$t$,
\begin{equation}
\langle\Delta r^{2}(t)\rangle
\equiv
\langle|{\bf r}(t)-{\bf r}(0)|^{2} \rangle
=
2\int_{0}^{t}{\rm d}s\,(t-s)R(s)\:,
\label{e:taylor}
\end{equation}
\noindent{}a result that assumes only steady state~\cite{Taylor1922,Majda1999}. The superdiffusive case of interest here corresponds to a divergent $\lim_{t\rightarrow\infty}\int^{t}{\rm d}s\,R(s)$, {\em i.e.} $R(t)\sim t^{-b}$ with $b<1$ (we ignore the marginal case $b=1$). According to (\ref{e:taylor}), this corresponds to $\langle\Delta r^{2}(t)\rangle\sim t^{1+a}$ with $a=1-b$. Thus superdiffusion corresponds to an anomalous slow decay of the velocity autocorrelation function.

\begin{table}
\caption{\label{t:exps}Exponents $a$, $b$, $\nu$, $\alpha$ and $\beta$ from power law fits for $\langle\Delta r^{2}(t)\rangle$ (MSD), collapse of $P(\Delta r^{2},t)$ (PDF), persistent directed motion~(PDM; eqn.~(\ref{e:pdm})), giant number fluctuations $\delta N/N^{1/2}$ (GNF) and static structure factor $S(q)$. $L=150\langle d\rangle$ and the numbers in brackets gives the error in the last digit.
}
\begin{ruledtabular}
\begin{tabular}{c@{\quad}|@{\quad}lllll}
Parameters
&
MSD\footnote{Fits to $\langle\Delta r^{2}(t)\rangle=[C/t^{1+a}+E/t]^{-1}$ for large $t$.}
&
PDF
&
PDM
&
GNF
&
$S(q)$\footnote{Fits to $S(q)=Fq^{-\beta}+G$ up to $q$ corresponding to $\approx$ 1---3$\langle d\rangle$.}
\\
 & $\sim t^{1+a}$ & $\sim t^{\nu}$ &$\sim t^{1-b}$ & $\sim N^{\alpha}$ & $\sim q^{-\beta}$ \\
\hline
\hline
$K/K_{0}=5\times10^{-3}$ & & & & & \\
\hline
$\eta/\eta_{0}=0.04$ & 0.4(1) & 1.35(5) & 0.4(1) & 0.2(1)\footnote{Not converged with system size (exponent increasing with $L$)} & 1.75(5) \\
$\eta/\eta_{0}=0.08$ & 0.60(5) & 1.35(5) & 0.5(1) & 0.3(1)$^{\rm c}$ & 1.75(5) \\
$\eta/\eta_{0}=0.24$ & 0.60(5) & 1.3(1) & 0.5(1) & 0.3(1)$^{\rm c}$ & 1.7(1) \\
\hline
\hline
$K/K_{0}=5\times10^{-4}$ & & & & & \\
\hline
$\eta/\eta_{0}=0.04$ & 0.45(5) & 1.45(5) & 0.5(1) & 0.2(1)$^{\rm c}$ & 2.00(5) \\
$\eta/\eta_{0}=0.08$ & 0.55(5) & 1.45(5) & 0.50(5) & 0.3(1)$^{\rm c}$ & 1.9(1) \\
$\eta/\eta_{0}=0.24$ & 0.65(5) & 1.40(5) & 0.6(1) & 0.3(1) & 1.7(1) \\
\hline
\hline
$K/K_{0}=5\times10^{-5}$ & & & & & \\
\hline
$\eta/\eta_{0}=0.04$ & 0.45(5) & 1.50(5) & 0.5(2) & 0.2(1)$^{\rm c}$ & 2.0(1) \\
$\eta/\eta_{0}=0.08$ & 0.50(5) & 1.50(5) & 0.5(1) & 0.3(1)$^{\rm c}$ & 1.8(2) \\
$\eta/\eta_{0}=0.24$\footnote{Used larger system size $L=200\langle d\rangle$.} & 0.65(5) & 1.50(5) & 0.6(1) & 0.3(1) & 1.8(1) \\
\end{tabular}
\end{ruledtabular}
\end{table}

For our 2--dimensional system, the value $b=\frac{1}{2}$ is expected based on simple scaling arguments as explained in~\cite{Fiege2009}. In brief, over a time scale $\tau$, the random force noise changes the momentum of each particle $\alpha$ by an amount $\Delta{\bf p}^{\alpha}\sim\tau^{1/2}$. These momentum fluctuations are smoothed by the short--range repulsive interaction and, since momentum is conserved, can only spread diffusively. Therefore on a time scale $\tau$ the momentum fluctuation $\Delta{\bf p}^{\alpha}$ will be smoothed over a volume $\sim\tau^{D/2}$ in $D$ dimensions. Thus momentum fluctuations should scale as $\Delta{\bf p}^{\alpha}$ divided by the number of particles over which the momentum has been shared, {\em i.e.} $\sim\tau^{1/2}/\tau^{D/2}\sim\tau^{-1/2}$ when $D=2$, so $R(t)\sim t^{-1/2}$ and $b=a=\frac{1}{2}$ as claimed.

\begin{figure}[htpb]
\center{\includegraphics[width=9cm]{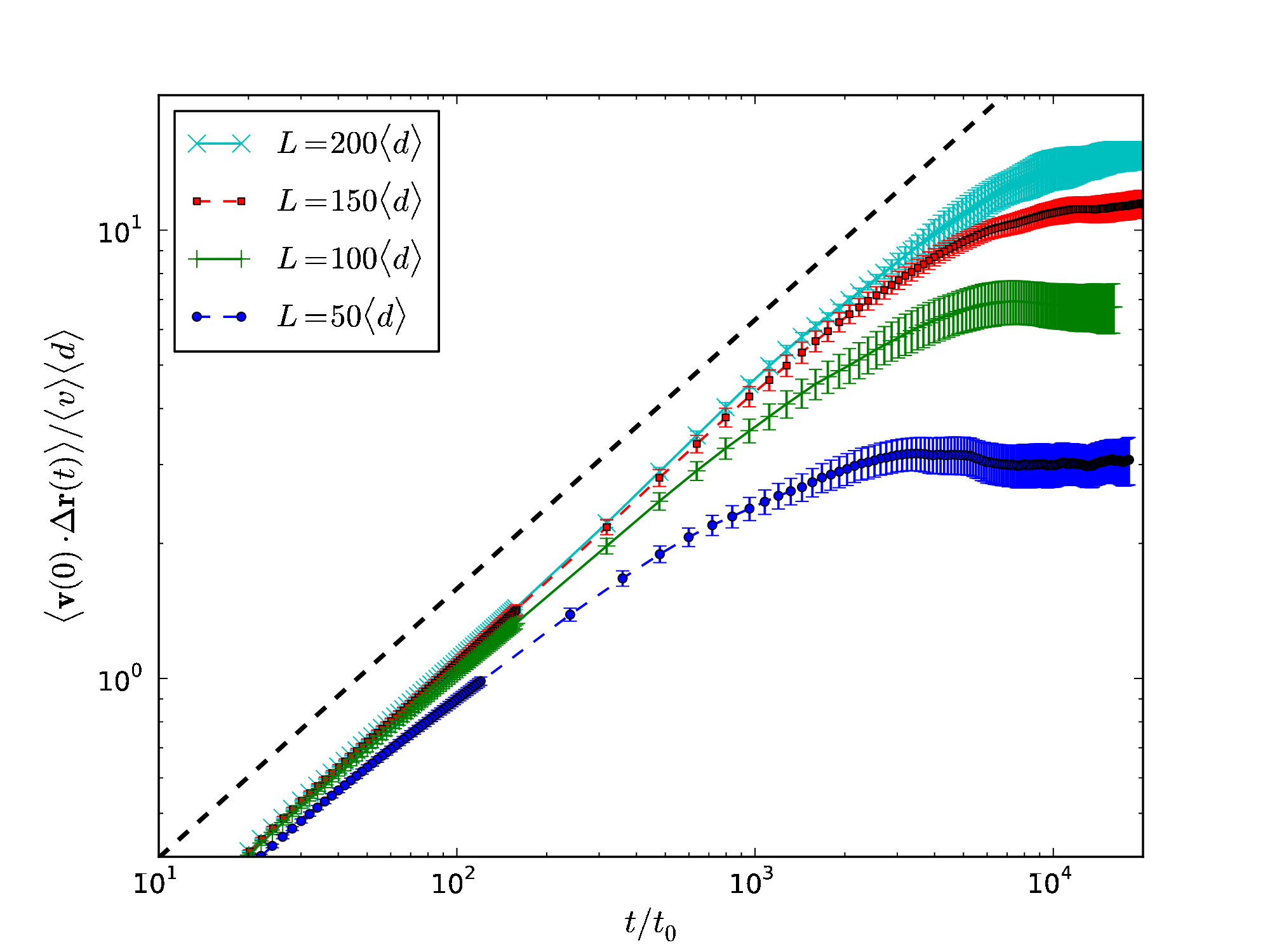}}
\caption{{\em (Color online)} $\langle {\bf v}(0)\cdot\Delta{\bf r}(t)\rangle$ normalized by $\langle d\rangle$ and the mean velocity $\langle v\rangle$ for the system sizes~$L$ given in the key, $K/K_{0}=5\times10^{-5}$ and $\eta/\eta_{0}=0.24$. The dashed line has a slope of~$0.6$.}
\label{f:pdm}
\end{figure}

A direct corollary of the slow decay of $R(t)$ is that particles drift over arbitrarily long distances. Put precisely, for $b<1$ the integral of $R(t)=\langle{\bf v}(0)\cdot{\bf v}(t)\rangle\sim t^{-b}$ with respect to the lag time~$t$ is unbounded, {\em i.e.}
\begin{equation}
\int_{0}^{t}{\rm d}t\,R(t)
=
\langle {\bf v}(0)\cdot\Delta{\bf r}(t)\rangle
\sim
t^{1-b}
\:,
\quad b<1\:,
\label{e:pdm}
\end{equation}
\noindent{}with $\Delta{\bf r}(t)={\bf r}(t)-{\bf r}(0)$ as before. Thus particles will tend to move arbitrarily far in the direction of their initial motion, in the statistical sense of (\ref{e:pdm}). We refer to this as {\em persistent directed motion}. We have directly measured the integral (\ref{e:pdm}) in our simulations and in all cases found power law growth, as demonstrated in Fig.~\ref{f:pdm}. Fitted values of the exponent are given in Table~\ref{t:exps} and are in all cases consistent with the predicted value \mbox{$1-b=1/2=a$}.


%
%

%
%
\begin{figure}[htpb]
\center{
\includegraphics[width=4.2cm]{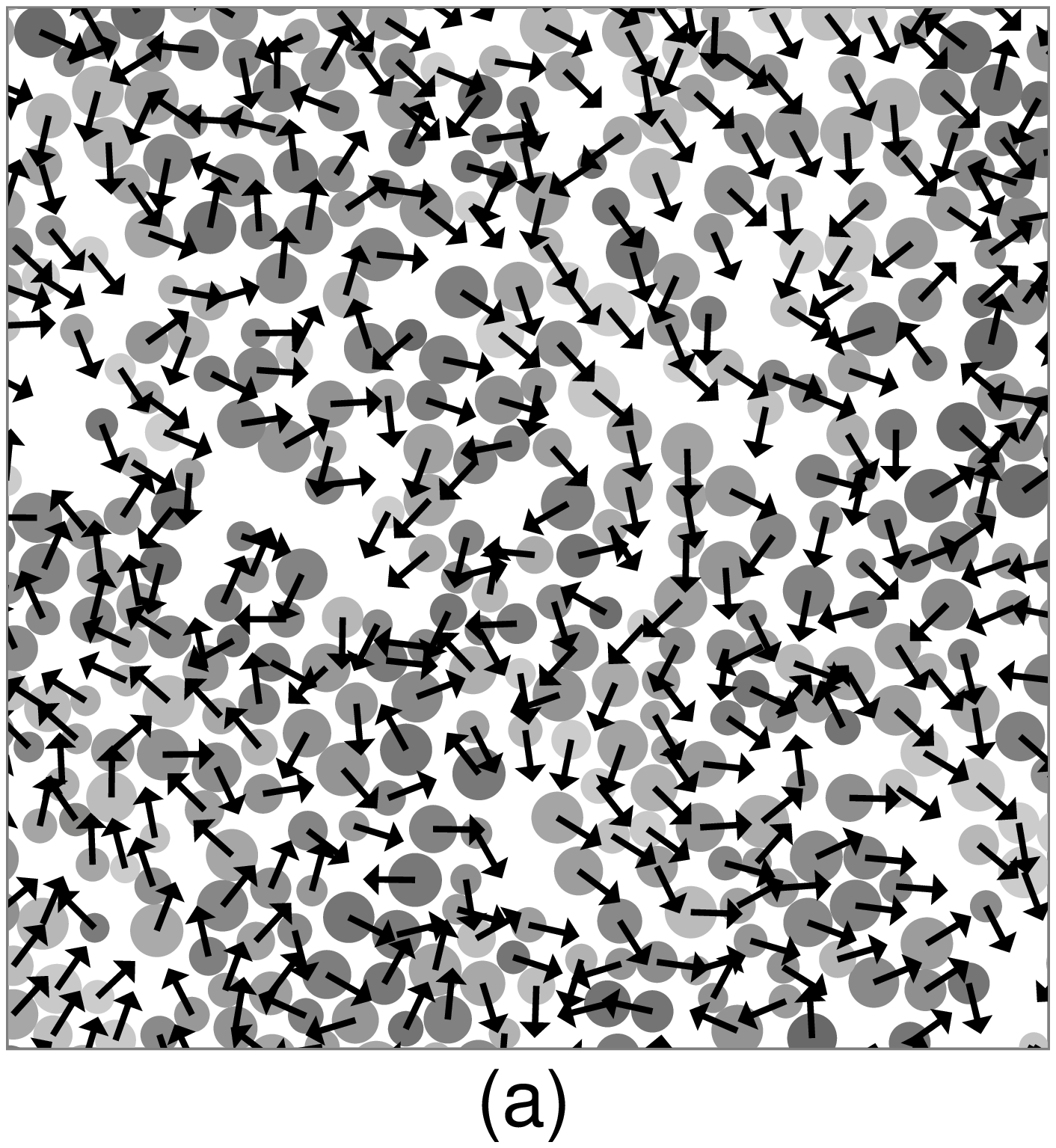}
\hfill
\includegraphics[width=4.2cm]{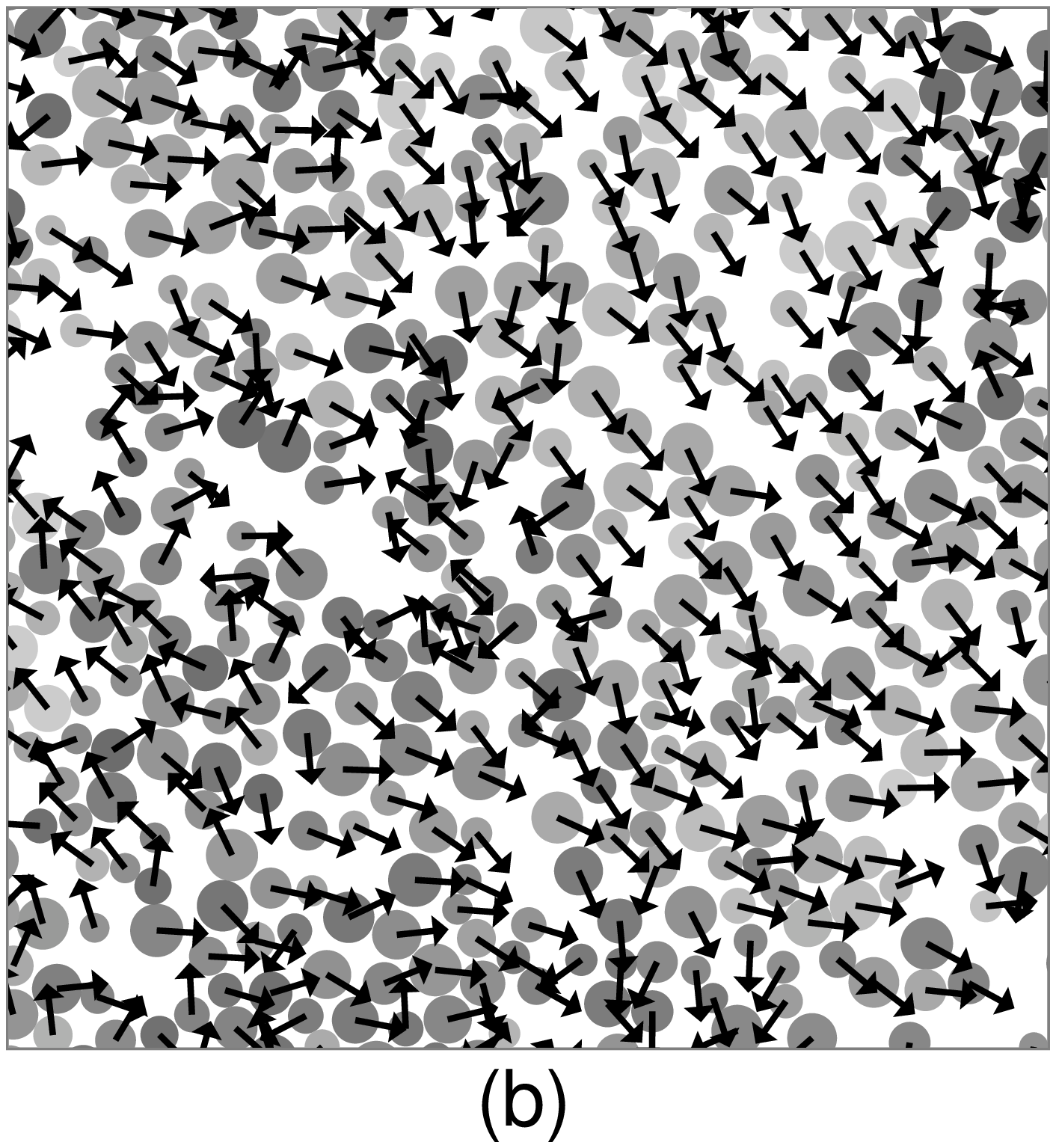}
}
\caption{{\em (a)}~Particle displacements over a time interval \mbox{$t/t_{0}\approx120$} for the same parameters as Fig.~\ref{f:pdm}, where arrows denote direction of total displacement over this interval, and light (dark) discs correspond to large (small) displacements. {\em (b)}~Same starting configuration for \mbox{$t/t_{0}\approx1200$}. The mean displacement is $\approx1.7\langle d\rangle$ in {\em (a)} and $\approx10.5\langle d\rangle$ in~{\em (b)}. Larger, color figures available from~\cite{EPAPS}.
}
\label{f:pdm_snapshots}
\end{figure}

Snapshots reveal that this single--particle quantity has a corresponding spatial signature. As evident in Fig.~\ref{f:pdm_snapshots}, particle motion becomes correlated over larger distances when longer time intervals are considered. To quantify this effect, note that according to~(\ref{e:pdm}), particles will move on average a distance $\delta r\sim\tau^{1-b}$ on a time scale~$\tau$. Using the same diffusive dispersion relation for momentum fluctuations as above, over the same time scale particle motion will become correlated over a range $\ell\sim\tau^{1/2}$. Thus of the $N\sim\ell^{D}$ particles in a region of size~$\ell$, a fraction $\delta N/N\sim\delta r/\ell$ will leave or enter the region, generating number fluctuations of magnitude
\begin{equation}
\delta N\sim\ell^{1-2b+D}\sim N^{\frac{1-2b}{D}+1}\:.
\label{e:dN}
\end{equation}
\noindent{}As long as (\ref{e:pdm}) is not subject to a finite size cut--off, persistent directed motion will apply over arbitrarily large length and time scales. Thus for all region sizes~$\ell$, there will be a corresponding time scale $\tau\sim\ell^{2}$ driving the fluctuations in~(\ref{e:dN}). Note that although inspection of snapshots suggests persistent motion breaks down primarily in low--density regions, such velocity--density couplings are not included in this simple theory.

\begin{figure}[htpb]
\center{\includegraphics[width=9cm]{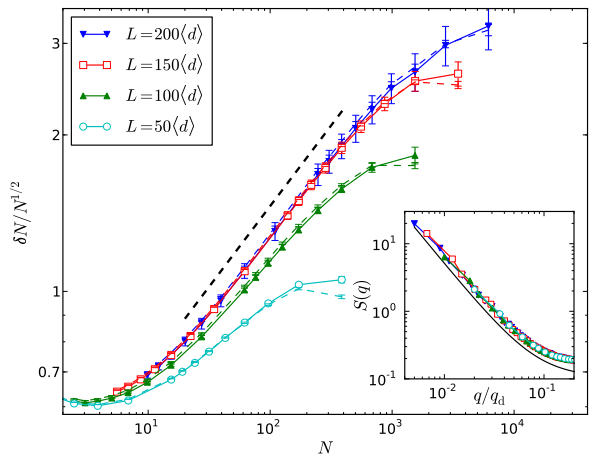}}
\caption{{\em (Color online)} Number fluctuations $\delta N/N^{1/2}$ versus $N$ for different system sizes $L$, and for the two different procedures P1 and P2 corresponding to time--averaging first (P1, solid lines) and spatial--averaging first (P2, dashed lines). $K/K_{0}=5\times10^{-5}$, $\eta/\eta_{0}=0.24$, and the dashed line has a slope of~0.31. {\em (Inset)}~Static structure factor $S(q)$ for the same systems. The solid line corresponds to $S(q)\propto q^{-1.8}+B$.
}
\label{f:gnf}
\end{figure}

These number fluctuations are `giant' when the exponent on the right--hand side of (\ref{e:dN}) is larger than~$1/2$. It is then straightforward to map fluctuations for large $N$ to a divergence at small wavelength $q$ in the static structure factor, $S(q)\sim q^{-[2(1-2b)+D]}$~\cite{GNF}. An example of both $\delta N/N^{1/2}$ and $S(q)$ is given in Fig.~\ref{f:gnf}, clearly demonstrating giant fluctuations and a divergent~\mbox{$S(q\rightarrow0)$}. The corresponding exponents for various parameters are given in Table~\ref{t:exps}, which should be compared to the prediction $\delta N/N^{1/2}\sim N^{1/2}$ and $S(q)\sim q^{-2}$ for $b=1/2$ and $D=2$. The number fluctuations are susceptible to finite size effects and we were only able to attain convergence for the 2 points marked in the table. To ensure convergence with time, $\delta N$ was measured following two procedures, one in which temporal averaging is performed before spatial averaging (P1), and a second in which the order of averaging is reversed (P2). These two measures agreed for all but the largest~$N$ as in Fig.~\ref{f:gnf}, indicating ergodicity~\cite{AransonNarayan} (we also checked that the intermediate scattering function decayed by at least an order of magnitude over the same interval). Where available the fitted exponents are not inconsistent with the prediction, and we note the slope of the curves monotonically increases with system size in all cases, so giant fluctuations will become more pronounced, not less, for infinite systems.

For $S(q)$ the picture is clearer. Given a similar magnitude of systematic error induced by the narrow fitting regime as before, the $S(q)$ exponents in Table~\ref{t:exps} are within reasonable distance of the prediction. A diverging $S(q)$ had also been observed in randomly driven inelastic hard--sphere systems, with the exponent $2$ predicted by granular hydrodynamics~\cite{GranSq,Ohta}, suggesting its existence may not depend on the details of the dissipation mechanism, {\em i.e.} whether it is scalar or vector. Super--ballistic mass transport was also claimed in hard--sphere simulations~\cite{Ohta}, but no causal link with $S(q)$ suggested.


%
%

%
%
We have thus demonstrated the co--existence of superdiffusion and large--scale structure in a driven dissipative system, and provided a simple theory implying a causal link between the two which agrees with the available numerical data. We do not claim this relationship is general, and cite the contrary examples of turbulence in incompressible fluids, which can be superdiffusive without density fluctuations~\cite{Majda1999}, and isochoric critical fluids, which have a divergent $S(q)$ but no superdiffusion~\cite{CritPt}. It is possible that superdiffusion inevitably leads to some form of long--range static or dynamic correlations that need not take the form of density fluctuations as in our model, but to confirm this would require more careful inspection of candidate systems. Additionally, our scaling theory does not include $K$ or~$\eta$; empirically we find a non--trivial dependency of the various prefactors on these quantities which cannot be decomposed into separate power laws for each variable. An enhanced theory predicting the full scaling laws would be desirable, as would any attempt to deepen our understanding of this potentially far--reaching non--equilibrium phenomenon.


%
%

%
%

\end{document}